# 6-O-glucose palmitate synthesis with lipase: Investigation of some key parameters

Dounia Arcens[a,b], Etienne Grau[a,b], Stéphane Grelier[a,b], Henri Cramail*[a,b] and Frédéric Peruch*[a,b]

[a] Univ. Bordeaux, CNRS, Bordeaux INP/ENSCBP, Laboratoire de Chimie des Polymères Organiques, UMR 5629, 16 avenue Pey-Berland, F-33607 Pessac Cedex, France

[b] Centre National de la Recherche Scientifique, Laboratoire de Chimie des Polymères Organiques, UMR 5629, 16 avenue Pey-Berland, F-33607 Pessac Cedex, France

**Abstract**

Fatty acid sugar esters represent an important class of non-ionic bio-based surfactants. They can be synthesized from vinyl fatty acids and sugars with enzyme as a catalyst. Herein, the influence of the solvent, the lipase and the temperature on a model reaction between vinyl palmitate and glucose *via* enzymatic catalysis has been investigated and the reaction conditions optimized. Full conversion into 6-O-glucose palmitate was reached in 40 hours in acetonitrile starting from a reactant ratio 1:1, at only 5 %-wt loading of lipase from *Candida antarctica* B (CALB) without the presence of molecular sieves.

**Funding**

This work was performed in partnership with the SAS PIVERT, within the frame of the French Institute for the Energy Transition (Institut pour la Transition Energétique - ITE) P.I.V.E.R.T. (www.institut-pivert.com) selected as an Investment for the Future ("Investissements d'Avenir"). This work was supported, as part of the Investments for the Future, by the French Government under the reference ANR-001-01.

# 1. Introduction

Fatty acid sugar esters are non-ionic surfactants that can be synthesized from inexpensive natural resources. Because of their amphiphilic nature, non-toxicity and biodegradability, [1,2] they find a wide range of applications in many fields such as food, [3] pharmaceutical, [4] detergents and cosmetics. [5] Depending on the chosen carbohydrate and acyl moieties, fatty acid sugar esters have been shown to exhibit anti-oxidant, [6] antimicrobial, [7-9] insecticidal, [10] and antitumoral [11] properties. They can be obtained by a chemical route using alkaline catalysts, [12] but this strategy requires high temperatures and the use of hazardous solvents such as DMF or pyridine, which are not compatible with food applications. Besides, as all the carbohydrate hydroxyl groups exhibit similar reactivity, it usually results in mixtures of esters, without any control of the composition. [13] Enzymes such as lipases, proteases and esterases are also able to catalyze fatty acid sugar ester synthesis with high selectivity, directly yielding mono-esters without need of additional protection/deprotection steps. Among them, lipases are the most used enzymes to catalyze fatty acid sugar esters synthesis. These enzymes are active in many organic solvents and at lower temperatures. Enzymatic route has therefore been widely studied as a milder and greener alternative to synthesize fatty acid sugar esters, but it presents some drawbacks such as longer reaction times, lower yields and an important cost, as large quantities of lipase are usually required (around 20 wt.%). Another major issue is also to find an appropriate solvent that can both solubilize the carbohydrate and the fatty acid moieties, without deactivating the lipases. Hydrophobic solvents enhance lipases activity, [14] but poorly solubilize carbohydrates. Tertiary alcohols such as *tert*-butanol, [15] and 2-methyl-butan-2-ol [16] are generally used as their relative polarity enables a good solubility of carbohydrates. Mixtures of two solvents such as *tert*- butanol/pyridine [17] or 2-methyl-butan-2-ol/DMSO [6, 18] have also been tested in order to increase the carbohydrate solubility. More recently, ionic liquids [19-22] have also been explored. Another challenge is to increase the final conversions into fatty acid sugar esters. Esterification leads to the formation of water, which must be removed to shift the equilibrium toward fatty acid sugar ester formation, for instance by adding molecular sieves to the reaction media. [23-26] Ducret *et al.* developed a process of fatty acid sugar esters synthesis under reduced pressure to remove water. [27] Another strategy is to start from fatty acid vinyl esters. In that case, the transesterification sub-product is acetaldehyde, which is easily removed, leading to fast and high conversions. [18, 22, 28-33] In the present work, 6-O-glucose palmitate was synthesized in classical organic solvents from a 1:1 ratio of glucose and vinyl palmitate mainly with Lipase B from *Candida antarctica* (CALB) as the catalyst. Only 5 %-wt of the supported lipase were used and the influence of the solvent, reaction time and presence of molecular sieves investigated. Several commercially available lipases were compared and the influence of the reaction temperature was also examined.

**2. Materials and methods**

*2.1 Materials*

Vinyl palmitate was purchased from TCI Europe and was dried under dynamic vacuum overnight prior to use. Anhydrous glucose was purchased from Fluka and lipase B from *Candida antarctica* supported on acrylic beads (activity > 5000 U/g) was purchased from Sigma Aldrich. Supported lipases IMMCALB-T2-150 from *Candida antarctica* B (2500 U/g), IMMCALA-T2-150 from *Candida antarctica* A (3000 U/g), IMMRML-T2-150 from *Rhizomucor miehei* (1500 U/g), IMMTLL-T2-150 f*rom Thermomyces lanuginosa* (10000 U/g), IMMABC-T2-150 from *Pseudomonas cepacia* (1500 U/g) and IMML51-T2-150 from *Fusarium solani pisi* (5000 U/g) were purchased from Chiral Vision. All lipases were used as received. Acetonitrile, THF, DMF, DMSO, and cyclohexane were purchased from Fluka, HPLC grade. Dichloromethane, HPLC grade and acetone, technical grade, were purchased from Sigma Aldrich. *tert*-Butanol, extra pure, was purchased from Acros Organics. Pyridine and dioxane were purchased from TCI. Solvent drying procedures are described in Supporting Information. Molecular sieves, 3Å, were purchased from Acros Organics and activated by heating at 400°C in a muffle furnace for 6 hours then flamed several times under dynamic vacuum. Once activated, the latter were stored in a glovebox. Deuterated DMSO was purchased from Eurisotop.

*2.2 General synthesis of 6-O-glucose palmitate catalyzed by CALB*

In a typical synthesis of 6-O-glucose palmitate, 0.9 mmol (249 mg) of vinyl palmitate and 0.9 mmol (162 mg) of glucose were poured into an oven-dried Schlenk with 10 mL of solvent under an argon flux. 20 mg of supported CALB were then added. When needed, 100 mg of activated 3Å molecular sieves beads were added. The reaction was carried out during 72h, under magnetic stirring at 250 rpm and heated at 45°C by means of a thermoset oil bath. For kinetic studies, 0.2 mL samples were sampled out and analyzed by $^1$H NMR spectroscopy. Four different compounds were identified: glucose, vinyl palmitate, 6-O glucose palmitate and palmitic acid. In each case, the primary alcohol of glucose was only esterified; all the secondary alcohols remaining untouched. At the end of the reaction, the solvent was evaporated. THF was poured in the crude mixture and the obtained mixture was filtered under vacuum in order to remove the lipase and most of the glucose. The soluble part was evaporated. The obtained solid was dispersed into water then filtrated on a Büchner in order to remove traces of glucose. 5-10 mL of acetone was then added to dissolve the unreacted fatty chains and the suspension was filtrated again. The remaining insoluble white powder was characterized by $^1$H NMR spectroscopy and was found to be pure 6-O-glucose palmitate. No significant loss was observed during the purification and 6-O-glucose palmitate was obtained with a yield of 90%. $^1$H NMR (*d6*-DMSO, 400 MHz, δ (ppm)): 0.8 (3H, t, CH$_3$), 1.1-1.2 (H, m, alkyl chain CH$_2$), 1.4 (2H, q, CH$_2$CH$_2$CO), 2.3 (2H, t, CH$_2$CO), 3.1 (1H, m, H4), 3.2 (1H, m, H2), 3.4 (1H, m, H3), 3.7 (1H, m, H5), 4.0 (1H, m, H6a), 4.25 (1H, m, H6b), 4.6 (1H, d, OH3), 4.7 (1H, d, OH2), 4.9 (1H, t, H1), 5.0 (d, OH4β), 5.1 (d, OH4α), 6.2 (d, OH1β), 6.55 (d, OH1α)

*2.3 Analysis*

2.3.1 NMR spectroscopy

NMR experiments were performed at 298K on a Bruker Avance 400 spectrometer operating at 400MHz. Deuterated DMSO was used as solvent.

2.3.2 HPLC

HPLC analysis were performed on a HPLC apparatus with an evaporating light scattering detector (ELSD, Varian 380-LC) and a Prevail carbohydrate ES 5μ column. The evaporator and nebulizer temperatures were set at 90°C and 40°C, respectively. 50 μL of the samples were injected. The eluent was a solution of 75/25/5 v/v/v methanol/acetonitrile/water with a flow rate of 0.5 mL.min$^{-1}$.

## 3. Results and discussion

Enzymatic fatty acid sugar ester synthesis is a complex process where several reactions can take place. CALB can catalyze vinyl palmitate transesterification into 6-O-glucose palmitate (reaction 1, Figure 1). Nevertheless, because of the presence of residual water in the reaction medium, the enzyme is also able to catalyze vinyl palmitate hydrolysis (reaction 2, Figure 1). Those two reactions are irreversible.

Besides, an equilibrium takes place between the so-formed palmitic acid and 6-O-glucose palmitate (reaction 3, Figure 1).

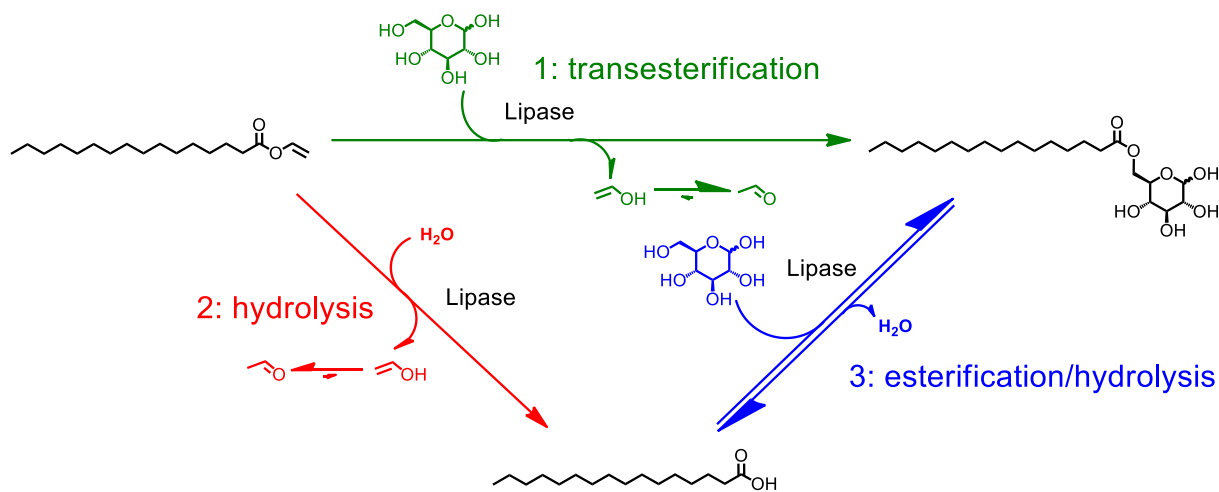

*Figure 1 – Reaction scheme between vinyl palmitate (VP) or palmitic acid (PA) with glucose to produce 6-O-glucose palmitate (GP) in the presence of CALB*

*3.1 Effect of the solvent*

Reaction between vinyl palmitate and glucose was carried out in ten organic solvents in the presence of CALB. Each solvent was tested as received (Conditions A) and also after drying (Conditions B). In the cases of acetone, acetonitrile, THF, *tert*-butanol and dioxane, the influence of the presence of 3Å molecular sieves beads in the reaction medium was investigated and kinetic studies of the corresponding reactions were performed (Conditions C). Conversions into 6-O-glucose palmitate after 72 hours of reaction are given in Table 1. In Figure 2, simultaneous variations of vinyl palmitate, palmitic acid and 6-O-glucose palmitate contents with time are shown for each solvent, without and in the presence of molecular sieves. From those plots, the initial reaction rates could be calculated and are given in Table 2. Among all the tested solvents, only five allowed glucose acylation: acetone, tert-butanol, tetrahydrofuran, dioxane, and acetonitrile. Cyclohexane gave no conversion into 6-O-glucose palmitate. This could be explained by a too poor glucose solubility in this solvent (see Table 3) but no hydrolysis was observed either, evocating the absence of activity of CALB. Yet CALB was reported to exhibit higher activity in apolar solvents, [34] and examples of transesterifications in hexane in the presence of CALB have been reported in literature. [34, 35] Same results were obtained in dichloromethane, DMF, DMSO and pyridine. CALB has been reported to be inactive in these solvents. [36] This was confirmed in our experiments as vinyl palmitate remained intact. If CALB was active, hydrolysis would have been observed.

| Solvent | Conditions A | Conditions B | Conditions C |
|---|---|---|---|
| Acetone | 12% [a] | 93% [a] | 100% |
| *tert*-Butanol | 32% [a] | 88% [a] | 94% [a] |
| THF | 52% [a] | 88% [a] | 100% |
| Dioxane | 49% [a] | 80% [a] | 90% [a] |
| Acetonitrile | 52% [a] | 100% | 100% |
| DCM | 0% [b] | 0% [b] | - |
| DMSO | 0% [b] | 0% [b] | - |
| DMF | 0% [b] | - | - |
| Cyclohexane | 0% [b] | 0% [b] | - |
| Pyridine | 0% [b] | - | - |

*Table 1 – Conversion of vinyl palmitate into 6-O-glucose palmitate in various organic solvents, after 72h at 45°C, determined by $^1$H NMR spectroscopy. MS = 3Å molecular sieves beads. (a) Vinyl palmitate was entirely converted into a mixture of 6-O-glucose palmitate and palmitic acid. (b) Vinyl palmitate remained intact: neither trans-esterification nor hydrolysis was observed*

As indicated in Table 1, the esterification reactions in solvents as received led to 6-O-glucose palmitate with 12% yield in acetone, 32% in tert-butanol, 49% in dioxane and 52% in acetonitrile and THF. As a general trend, drying the solvents enables to increase the conversions: 80% in dioxane, 88% in *tert*-butanol and THF, 93% in acetone and full conversion in acetonitrile. The addition of 3Å molecular sieves beads in the reaction media enabled the conversions in acetone and THF to reach 100%. These results highlight the influence of residual water content in the reaction medium on the conversions into glycolipids: the use of anhydrous solvents is thus necessary to obtain high conversions.

| Solvent | Conditions B | | | | Conditions C | | | |
|---|---|---|---|---|---|---|---|---|
| | $k_{VP}$ | $k_{PA}$ | $k_{GL}$ | $k_{GL}/k_{PA}$ | $k_{VP}$ | $k_{PA}$ | $k_{GL}$ | $k_{GL}/k_{PA}$ |
| Acetone | -91 | 17 | 74 | 4.3 | -103 | 10 | 93 | 9.3 |
| Acetonitrile | -69 | 32 | 56 | 1.7 | -69 | 22 | 57 | 2.6 |
| *tert*-butanol | -75 | 42 | 35 | 0.8 | -53 | 20 | 31 | 1.5 |
| Dioxane | -78 | 22 | 55 | 2.5 | -92 | 33 | 59 | 1.8 |
| tetrahydrofuran | -81 | 42 | 40 | 0.9 | -63 | 15 | 54 | 3.5 |

*Table 2 – Estimation of initial rates ($\mu mol.min^{-1}.g^{-1}$ of lipase) of vinyl palmitate (VP) consumption, palmitic acid (PA) and 6-O-glucose palmitate (GL) formation with a 10 % error. B = without molecular sieves; C = in the presence of 3Å activated molecular sieves beads (10mg per mL of solvent).*

The reaction process can be monitored by $^1$H NMR analysis by sampling out aliquots along with the reaction in order to follow conversions with time of vinyl palmitate, palmitic acid and 6-O-glucose palmitate, respectively. Glycolipid formation was shown to be the fastest in acetone, reaching the rate of 74 $\mu$mol.min$^{-1}$/g of lipase. When adding molecular sieves to the reaction medium, the reaction rate increases to 93 $\mu$mol.min$^{-1}$.g$^{-1}$ while it does not exceed 60 $\mu$mol.min$^{-1}$.g$^{-1}$ in all other tested solvents. Moreover, percentage of palmitic acid in the reaction medium is significantly lower in acetone than in the other solvents; it does not exceed 20% while it reaches 53% in *tert*-butanol (see Figure 2). In acetone, glycolipid formation seems to be more favorable than hydrolysis as indicated by higher $k_{GL}/k_{PA}$ in this solvent. Nevertheless, the reaction reaches an equilibrium in only 24 hours and requires the addition of molecular sieves to reach a full conversion into 6-O-glucose palmitate, whereas in acetonitrile, there is no need to add molecular sieves to reach full conversion as the formed palmitic acid was consumed and no equilibrium was observed.

Kinetic data (Figure 2) show that in all solvents excepted in dioxane, glucose acylation was still going on after vinyl palmitate disappearance as an esterification reaction takes place between the so-formed palmitic acid and the remaining glucose. In dioxane, this reaction does not occur and an equilibrium is achieved as soon as the vinyl palmitate vanishes. However, in the presence of molecular sieves, the removal of water enables the partial esterification of palmitic acid, finally increasing the conversion to 90%.

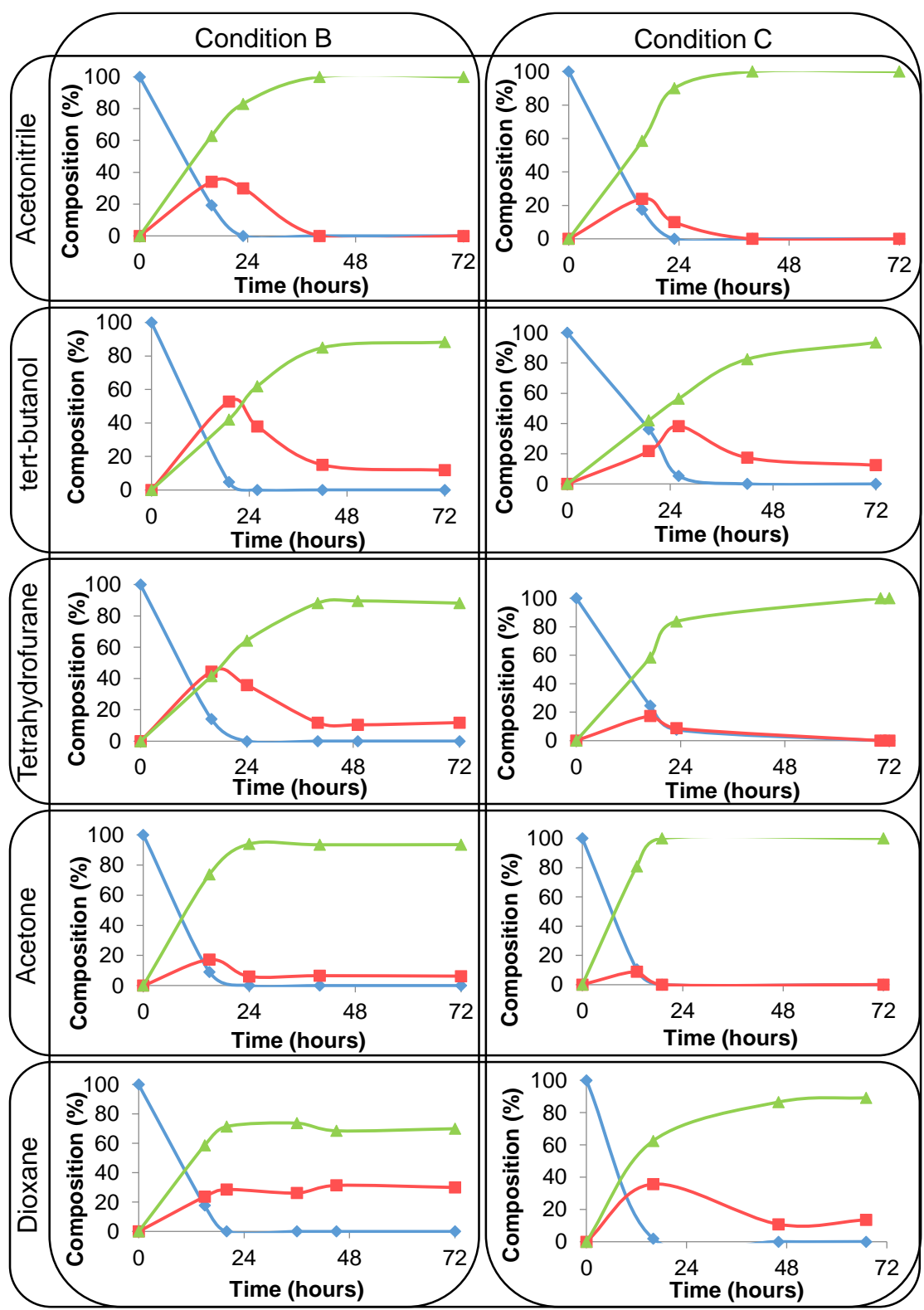

*Figure 2 – Composition in vinyl palmitate (blue), palmitic acid (red) and 6-O-glucose palmitate (green) as a function of time during the glycosylation of vinyl palmitate in acetonitrile, tert-butanol, THF, acetone and dioxane. B: without molecular sieves. C: in the presence of 3Å activated molecular sieves beads (10 mg/mL of solvent).*

| Solvent | Glucose solubility (mg/mL) | 6-O-Glucose palmitate solubility (mg/mL) |
|---|---|---|
| Acetone | 0.27 | 1.21 |
| Acetonitrile | 0.09 | 0.17 |
| *tert*-Butanol | 0.95 | 3.93 |
| Cyclohexane | ND[a] | - |
| Dichloromethane | ND[a] | - |
| Dimethylformamide | ND[b] | - |
| Dimethylsulfoxide | ND[b] | - |
| Dioxane | 0.70 | 2.45 |
| Pyridine | ND[b] | - |
| Tetrahydrofuran | 0.43 | 3.35 |

*Table 3 – Glucose and 6-O-glucose palmitate solubilities in the tested solvents. ND = not detected. [a] too low values to be measured; [b] glucose was fully soluble in the reaction medium*

These kinetic results can be explained by the solubility of the precursors and products. Because of its relatively high glucose solubility (0.95 mg/mL) in comparison to other organic solvents, *tert*-butanol is a good solvent candidate for glycolipid synthesis. [15] However, because of an important quantity of residual water even after 3 successive drying procedures (See supporting information), vinyl palmitate hydrolysis was by far the most important in *tert*-butanol: 53% of palmitic acid was reached, with an initial hydrolysis rate of 42 $\mu$mol.min$^{-1}$.g$^{-1}$ of lipase (Table 2). Palmitic acid can acylate glucose in *tert*-butanol, but much slower than vinyl palmitate, thus reaction kinetics were slow in *tert*-butanol. Similar hydrolysis rates are observed in THF, leading to lower palmitic acid amounts (42%) but both solvents lead to final conversions of 88%. The gap widened in presence of molecular sieves: 38% of palmitic acid is formed in *tert*- butanol at a rate of 20 μmol.min$^{-1}$.g$^{-1}$ of lipase, while only 17% are reached in THF at a rate of 15 μmol.min$^{-1}$.g$^{-1}$. This is probably why full conversion can be reached in 72h in THF in the presence of molecular sieves. Acetonitrile shows extremely low 6-O-glucose palmitate solubility, nearly 20 times lower than THF. As soon as 6-O-glucose palmitate is formed, it instantaneously precipitates out of the reaction medium to form a white solid crust at the surface. The equilibrium is therefore shifted into its formation. This probably explains why it is the only solvent allowing a full glucose acylation without molecular sieves. It is therefore the most suitable solvent of 6-O-glucose palmitate enzymatic synthesis.

*3.2 Enzymes*

In addition to the type of solvent, another important parameter is the catalyst used. Supported lipase from *Candida antarctica* (CALA and CALB), *Rhizomucor miehei* (RML), *Thermomyces lanuginosa* (TLL), *Pseudomonas cepacia* (ABC) and *Fusarium solani pisi* (L51) were used as catalysts for 6-O-glucose palmitate synthesis and the obtained conversions were compared. For these experiments, the amount of lipase to be used was calculated to correspond to 80 U. In the case of the lipase B from *Candida antarctica*, the influence of the linkage to the acrylic beads (adsorption or covalent linkage) was also investigated. All lipases were tested in 5 solvents: acetonitrile, acetone, *tert*-butanol, dioxane and THF. The esterification reactions were carried at 45°C under argon for 72 hours. The obtained conversions in 6-O-glucose palmitate are shown in Table 4. Whatever the solvent used, the best conversions were reached using supported CALB as a catalyst (either covalently-linked or absorbed on the support). Covalently-linked CALB led to lower conversions than adsorbed CALB, from 57% in tert-butanol to 85% in dioxane. It can be assumed that depending on which part of the enzyme is linked to the support, the active site can be more or less accessible by the substrates. The linkage could also partially distort the active site and therefore hinder the substrate complexation. Some glucose monoester was formed using the enzymes RML and TLL. Conversions up to 30% were observed in acetonitrile, acetone and tert-butanol with TLL. RML was less efficient: conversions of 18% in acetonitrile, 11% in dioxane and 8% in THF were respectively obtained, but this enzyme was not able to catalyze the esterification reaction in acetone and tert-butanol. CALA, ABC and L51 were found to be not efficient for the formation of glycolipid as only partial hydrolysis of vinyl palmitate was observed, except in acetone and in the presence of CALA, where no hydrolysis occurred either. The absence of glycolipid formation was thus not due to a deactivation of the enzyme but such lipases were not able to catalyze this esterification reaction. Glucose could be a bad substrate for these enzymes, maybe because of the geometry of their active site. TLL and RML exhibit crevice-like active sites, more accessible for the substrates. It is likely why they were able to catalyze the formation of glycolipid.

|  | Acetonitrile | Acetone | tert-Butanol | Dioxane | THF |
|---|---|---|---|---|---|
| *no enzyme* | 0[a] | 0[a] | 0[a] | 0[a] | 0[a] |
| *CALB adsorbed* | 100 | 93 | 88 | 80 | 88 |
| *CALB covalent* | 74 | 60 | 57 | 85 | 69 |
| *CALA* | 0[b] | 0[a] | 0[b] | 0[b] | 0[b] |
| *RML* | 18 | 2 | 3 | 11 | 8 |
| *TLL* | 36 | 28 | 32 | 6 | 9 |
| *ABC* | 1 | 0[b] | 3 | 9 | 4 |
| *L51* | 0[b] | 0[b] | 0[b] | 4 | 0[b] |

*Table 4 – Conversions of vinyl palmitate (VP) into 6-O-glucose palmitate in several solvents in the presence of various lipases. [a] only VP was observed: neither transesterification nor hydrolysis occurred; [b] hydrolysis into palmitic acid was observed.*

## 3.3 Temperature effect

The influence of temperature on the final conversion into 6-O-glucose palmitate and initial conversion rate was investigated for CALB catalysis in acetonitrile. Reactions were performed for 40 hours under argon at 20°C, 30°C, 45°C, 60°C and 70°C. All experiments were monitored by $^1$H NMR spectroscopy. Conversions versus time are plotted in Figure 3. Initial conversion rates were calculated based on conversions obtained from 0 to 8 hours for reactions at 60°C and 70°C and from 0 to 22 hours for lower temperatures. Data are collected in Table 5.

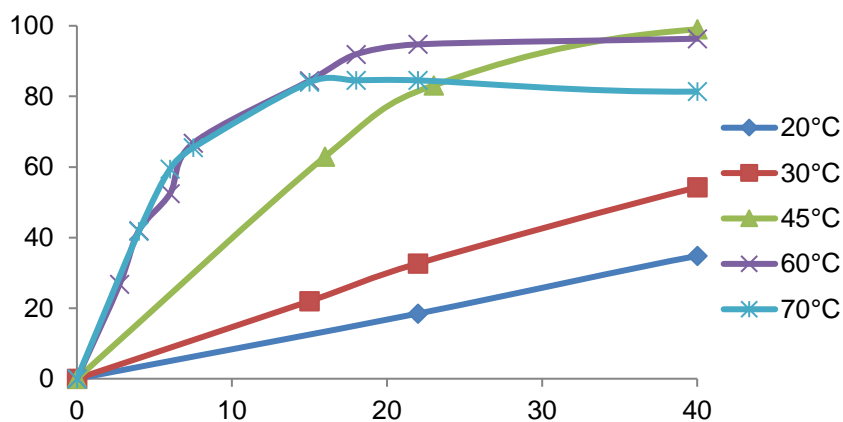

Figure 3 - Conversion into 6-O-glucose palmitate vs time in acetonitrile at different temperatures

| T (°C) | Initial conv. rate ($\mu$mol.min$^{-1}$.g$^{-1}$ lipase) | Conv. after 40h | Conv. after 67h |
|---|---|---|---|
| 20 | 6 | 35 | 49 |
| 30 | 10 | 54 | 73 |
| 45 | 28 | 100 | 100 |
| 60 | 68 | 96 | 96 |
| 70 | 70 | 81 | 86 |

*Table 5 – Influence of the temperature on the conversions into 6-O-glucose palmitate after 40h and 67h of reaction and the initial conversion rates*

From 20°C to 60°C, higher temperatures lead to faster conversions into 6-O-glucose palmitate. Heating at 60°C significantly reduces the reaction time: 94% of conversion were reached in only 20 hours, while it takes 40 hours to reach a full conversion at 45°C. Heating at 70°C did not increase reaction rates, but conversion reached a plateau at 86%, probably because of lipase deactivation.

## 4. Conclusion

The influence of the types of solvent and lipase as well as the reaction temperature on 6-O-glucose palmitate formation was investigated. Acetonitrile was shown to be the most suitable solvent, due to a very low glycolipid solubility: a complete glucose acylation was obtained in 40 hours in the presence of very low amounts of lipase. CALB gave the best conversions into glycolipid. Full glucose acylations were observed at 45°C and 60°C, but partial deactivation of the enzyme was observed at higher temperature.

## 5. Acknowledgements

The authors would like to thank Equipex Xyloforest ANR-10-EQPX-16 XYLOFOREST for HPLC.

**6-O-glucose palmitate synthesis with lipase: Investigation of some key parameters**

Dounia Arcens[a,b], Etienne Grau[a,b], Stéphane Grelier[a,b], Henri Cramail*[a,b] and Frédéric Peruch*[a,b]

[a] Univ. Bordeaux, CNRS, Bordeaux INP/ENSCBP, Laboratoire de Chimie des Polymères Organiques, UMR 5629, 16 avenue Pey-Berland, F-33607 Pessac Cedex, France

[b] Centre National de la Recherche Scientifique, Laboratoire de Chimie des Polymères Organiques, UMR 5629, 16 avenue Pey-Berland, F-33607 Pessac Cedex, France

**Supplementary information**

## 1. Solvent drying procedures

Cyclohexane, dimethylsulfoxide and dimethylformamide were dried through alumina columns and used immediately. The other dried solvents were distilled from the desiccant right before use.

| Solvent | 1st drying | 2nd drying | 3rd drying |
|---|---|---|---|
| Acetone | 72h on $CaSO_4$ | 72h on 3Å molecular sieves beads | - |
| Acetonitrile | 72h on $P_2O_5$ | 72h on 3Å molecular sieves beads | - |
| *tert*-Butanol | 72h on $CaSO_4$ | 24h on Na/benzophenone | 72h on 3Å molecular sieves beads |
| Cyclohexane | Dried through an alumina column | - | - |
| Dichloromethane | 72h on $CaH_2$ | - | - |
| Dimethylsulfoxide | Dried through an alumina column | - | - |
| Dimethylformamide | Dried through an alumina column | - | - |
| Dioxane | 72h on $CaH_2$ | - | - |
| Tetrahydrofuran | 72h on Na/benzophenone | 72h on 3Å molecular sieves beads | - |

*Table S1: Drying procedures applied for each solvent*

## 2. Determination of glucose and 6-O-glucose palmitate solubilities by HPLC

HPLC conditions are given in the Materials and Methods section

*2.1 Glucose*

2.1.1 Calibration curve

Calibration curve for glucose was determined by injecting in HPLC samples from six glucose solutions in water at known concentrations. For each sample, peak areas were measured and the obtained values were plotted versus injected mass. The relation between measured areas ($A_G$) and corresponding glucose mass ($m_G$) was obtained by linear regression.

| Sample | Concentration (mg/mL) | Injected volume (µL) | Corresponding mass (mg) | Measured area (pA*min) |
|---|---|---|---|---|
| 1 | 1.50 | 10 | 0.015 | 13.75 |
| 2 | 1.25 | 10 | 0.0125 | 11.82 |
| 3 | 1.00 | 10 | 0.01 | 9.62 |
| 4 | 0.75 | 10 | 0.0075 | 7.17 |
| 5 | 0.50 | 10 | 0.005 | 5.18 |
| 6 | 0.25 | 10 | 0.0025 | 2.54 |

*Table S2: Measured areas by HPLC for six glucose solution samples*

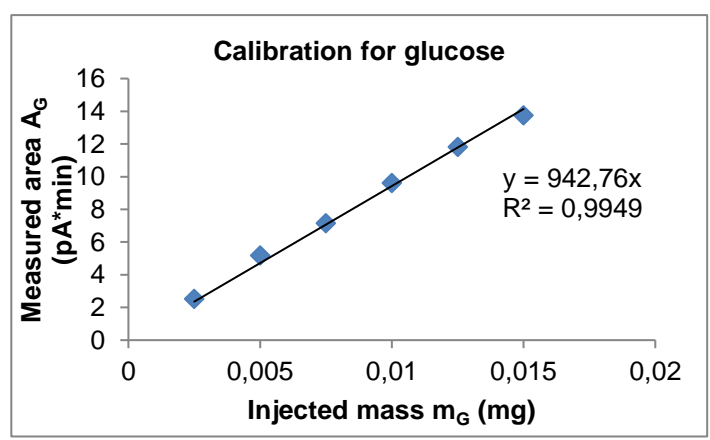

*Figure S1: Calibration curve for glucose*

$$A_G = 942.76 * m_G \quad (1)$$

2.1.2 Determination of glucose solubility in acetone, tert-butanol, acetonitrile, THF and dioxane

An excess of glucose was put in capped glass vials containing a magnetic bar, in presence of 1 mL of solvent. For each solvent, 2 or 3 samples (2 in acetone) were prepared. All the vials were put under agitation in a thermostated oil bath at 25°C for 72 hours. The samples were filtrated on 0.4 µm cellulose filters to remove insoluble glucose and the soluble parts were analyzed by HPLC. For each sample, the glucose peak areas were measured and the corresponding concentrations were calculated using Equation (1). All results are plotted in Table S3.

| Solvent | Sample | Injected volume (μL) | Measured area | Corresponding mass (mg) | Corresponding concentration (mg/mL) | Average concentration | |
|---|---|---|---|---|---|---|---|
| | | | | | | mg/mL | mM |
| Acetone | 1 | 50 | 12.78 | 0.0136 | 0.271 | 0.27 | 1.49 |
| | 2 | 50 | 12.55 | 0.0133 | 0.266 | | |
| t-BuOH | 1 | 50 | 45.27 | 0.0480 | 0.960 | 0.95 | 5.25 |
| | 2 | 50 | 43.89 | 0.0466 | 0.931 | | |
| | 3 | 50 | 44.52 | 0.0472 | 0.944 | | |
| Acetonitrile | 1 | 50 | 3.94 | 0.0042 | 0.084 | 0.09 | 0.50 |
| | 2 | 50 | 3.85 | 0.0041 | 0.082 | | |
| | 3 | 50 | 5.00 | 0.0053 | 0.106 | | |
| THF | 1 | 50 | 20.05 | 0.0213 | 0.425 | 0.43 | 2.40 |
| | 2 | 50 | 19.10 | 0.0203 | 0.405 | | |
| | 3 | 50 | 22.06 | 0.0234 | 0.468 | | |
| Dioxane | 1 | 50 | 33.73 | 0.0358 | 0.716 | 0.70 | 3.90 |
| | 2 | 50 | 32.12 | 0.0340 | 0.681 | | |
| | 3 | 50 | 33.48 | 0.0355 | 0.710 | | |

*Table S3: Determination of glucose solubility in various organic solvents*

## 2.2 6-O-glucose palmitate

Conditions are given in the Materials and Methods section

### 2.2.1 Calibration curve

Calibration curve for glucose was determined by injecting in HPLC samples from five 6-O-glucose palmitate solutions in DMSO at known concentrations. For each sample, peak areas were measured and the obtained values were plotted versus injected mass. The relation between measured areas ($A_{GP}$) and corresponding 6-O-glucose palmitate concentration ($C_{GP}$) was obtained by linear regression.

| Sample | Concentration (mg/mL) | Injected volume (µL) | Corresponding mass (mg) | Measured area (pA*min) |
|---|---|---|---|---|
| 1 | 10 | 50 | 0.5 | 62.04 |
| 2 | 5 | 50 | 0.25 | 31.07 |
| 3 | 1 | 50 | 0.05 | 15.00 |
| 4 | 0.5 | 50 | 0.025 | 9.89 |
| 5 | 0.25 | 50 | 0.0125 | 6.28 |

*Table S4: Measured areas by HPLC for 6 O-glucose palmitate solution samples*

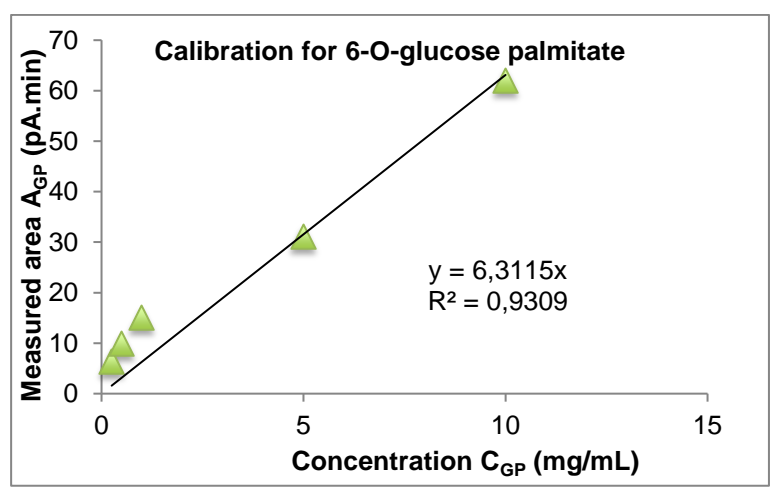

*Figure S2: Calibration curve for 6 O-glucose palmitate*

$$A_{GP} = 6.31 * C_{GP} \quad (2)$$

| Solvent | Injected volume (μL) | Measured area | Concentration | |
|---|---|---|---|---|
| | | | mg/mL | mM |
| Acetone | 50 | 7.662 | 1.21 | 2.9 |
| t-BuOH | 50 | 24.796 | 3.93 | 9.4 |
| Acetonitrile | 50 | 1.061 | 0.17 | 0.41 |
| THF | 50 | 21.143 | 3.35 | 0.80 |
| Dioxane | 50 | 15.461 | 2.45 | 0.59 |

*Table S5: Determination of 6 O-glucose palmitate solubility in various organic solvents*

## 3. Calculation of conversions into 6-O-glucose palmitate based on $^1$H NMR spectra:

$^1$H NMR spectra of vinyl palmitate, palmitic acid, 6-O-glucose palmitate, glucose and crude mixture after 16 hours of reaction are plotted in Figure 7. Attributions have been checked by COSY, HSQC and HMBC NMR. All signals and their attributions are plotted in Table 10. Vinyl palmitate, palmitic acid and 6-O-glucose palmitate contents have been calculated using Equation S1.

$$\%VP = \frac{\frac{1}{2}*(0.5\,I_{2.41\,ppm} + I_{7.21\,ppm})}{\frac{1}{2}*(0.5\,I_{2.41\,ppm} + I_{7.21\,ppm}) + \frac{1}{2}(0.5\,I_{2.17\,ppm} + I_{11.93\,ppm}) + \frac{1}{7}*(0.5\,I_{2.26\,ppm} + I_{3.76\,ppm} + I_{3.99\,ppm} + I_{5.01\,ppm} + I_{5.07\,ppm} + I_{6.33\,ppm} + I_{6.64\,ppm})}$$

$$\%PA = \frac{\frac{1}{2}(0.5\,I_{2.17\,ppm} + I_{11.93\,ppm})}{\frac{1}{2}*(0.5\,I_{2.41\,ppm} + I_{7.21\,ppm}) + \frac{1}{2}(0.5\,I_{2.17\,ppm} + I_{11.93\,ppm}) + \frac{1}{7}*(0.5\,I_{2.26\,ppm} + I_{3.76\,ppm} + I_{3.99\,ppm} + I_{5.01\,ppm} + I_{5.07\,ppm} + I_{6.33\,ppm} + I_{6.64\,ppm})}$$

$$\%GP = \frac{\frac{1}{7}*(0.5\,I_{2.26\,ppm} + I_{3.76\,ppm} + I_{3.99\,ppm} + I_{5.01\,ppm} + I_{5.07\,ppm} + I_{6.33\,ppm} + I_{6.64\,ppm})}{\frac{1}{2}*(0.5\,I_{2.41\,ppm} + I_{7.21\,ppm}) + \frac{1}{2}(0.5\,I_{2.17\,ppm} + I_{11.93\,ppm}) + \frac{1}{7}*(0.5\,I_{2.26\,ppm} + I_{3.76\,ppm} + I_{3.99\,ppm} + I_{5.01\,ppm} + I_{5.07\,ppm} + I_{6.33\,ppm} + I_{6.64\,ppm})}$$

*Equation S1: Calculation of vinyl palmitate (VP), palmitic acid (PA), and 6-O-glucose palmitate (GP) percentages of a given crude sample. I represents the integral value for each signal on the corresponding NMR spectrum. Integrals were calculated by setting $I_{0.85\,ppm}$ = 3*

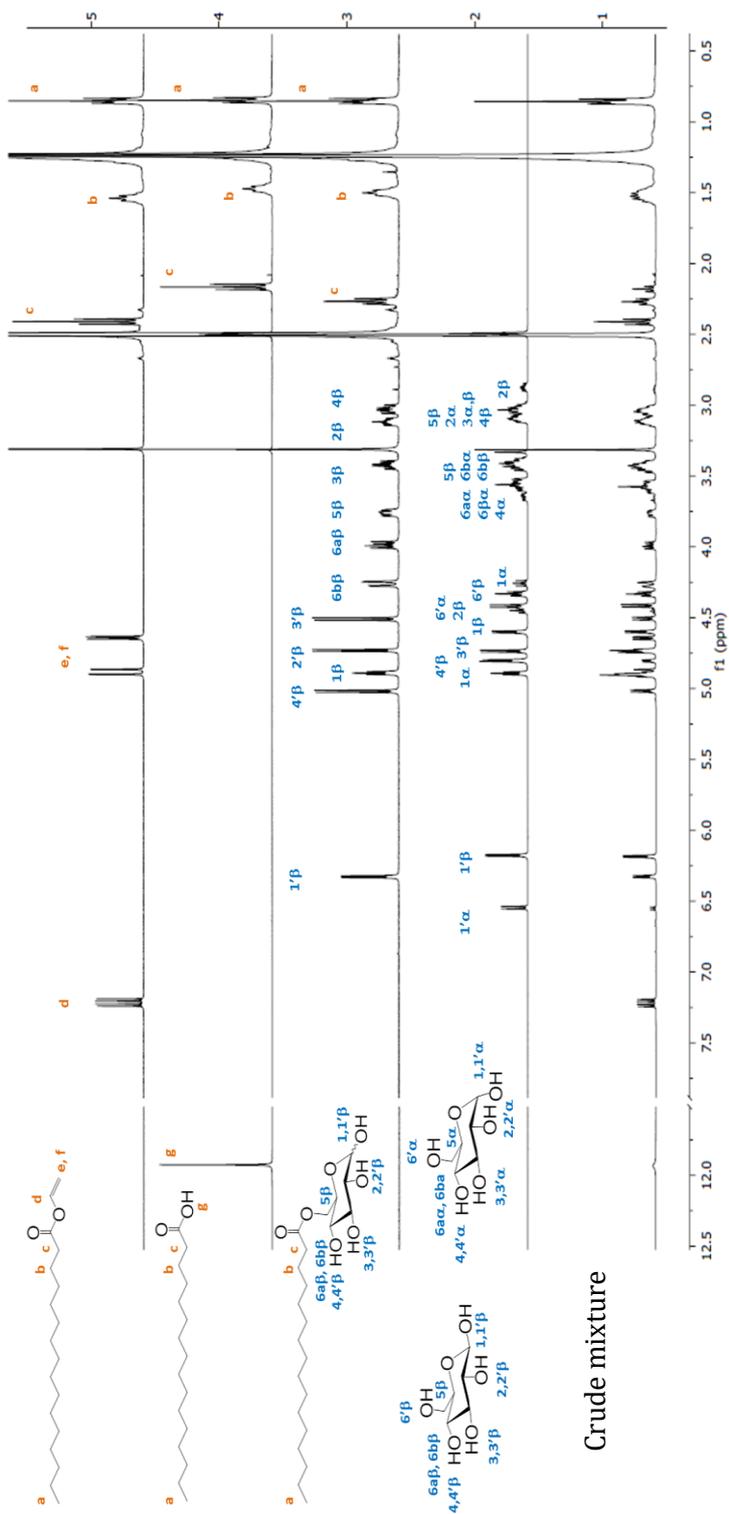

*Figure S3: Stacked $^1$H NMR spectra of (from top to bottom) pure vinyl palmitate, pure palmitic acid, pure 6-O-glucose palmitate, pure glucose, and crude reaction mixture after 16h (conditions: vinyl palmitate/glucose ratio = 1/1, concentration = 90mM, 5% 20 mg of CALB, in 10 mL of anhydrous acetonitrile; reaction performed at 45°C)*

*Table S6: List of chemical shifts on a typical raw mixture of vinyl palmitate, palmitic acid, glucose and 6-O-glucose palmitate*

| Chemical shifts (ppm) | Vinyl palmitate | Palmitic acid | Glucose | 6-O-glucose palmitate |
|---|---|---|---|---|
| 0.85 | $CH_3$ | $CH_3$ | | |
| 1.14-1.34 | $CH_2$ alkyl chain | $CH_2$ alkyl chain | | $CH_2$ alkyl chain |
| 1.54 | **$CH_2$**-CH2-CO | **$CH_2$**-CH2-CO | | **$CH_2$**-$CH_2$-CO |
| 2.17 | | **$CH_2$**-CO | | |
| 2.26 | | | | **$CH_2$**-CO |
| 2.41 | **$CH_2$**-CO | | | |
| 2.89 | | | H2β | |
| 3.04 | | | | H4 |
| 3.10 | | | | H2 |
| 3.42 | | | H5α, H6bα, H6bβ | H3 |
| 3.57 | | | H4α, H6aβ | |
| 3.66 | | | H6aα | |
| 3.76 | | | | H5 |
| 3.99 | | | | H6a |
| 4.26 | | | H1β | H6b |
| 4.33 | | | H6'β | |
| 4.41 | | | H2'β | |
| 4.45 | | | H6'α | |
| 4.50 | | | | |
| 4.56 | **$CH_2$**=CH | | | |
| 4.60 | | | H3'β | |
| 4.73 | | | | H3' |

| | | | |
|---|---|---|---|
| **4.74** | | H4'β | |
| **4.80** | | H2'α, H3'α, H4'α | |
| **4.86** | **CH$_2$**=CH | | |
| **4.90** | | H1α | H1 |
| **5.01** | | | H4'β |
| **5.07** | | | H4'α |
| **6.18** | | H1'β | |
| **6.33** | | | H1'β |
| **6.55** | | H1'α | |
| **6.64** | | | H1'α |
| **7.21** | **CH**=CH$_2$ | | |
| **11.93** | COOH | | |